\documentclass[sigconf]{acmart} 

\copyrightyear{2024} 
\acmYear{2024} 
\setcopyright{rightsretained} 
\acmConference[ASE '24]{39th IEEE/ACM International Conference on Automated Software Engineering }{October 27-November 1, 2024}{Sacramento, CA, USA}
\acmBooktitle{39th IEEE/ACM International Conference on Automated Software Engineering (ASE '24), October 27-November 1, 2024, Sacramento, CA, USA}
\acmDOI{10.1145/3691620.3695022}
\acmISBN{979-8-4007-1248-7/24/10}

\newif\ifDEBUG
\DEBUGfalse

\newif\ifBLINDED
\BLINDEDfalse

\newif\ifARXIV
\ARXIVfalse

\usepackage{tcolorbox}
\usepackage[final, commandnameprefix=ifneeded]{changes} 

\newcommand{\ie}{\textit{i.e.,\ }}
\newcommand{\eg}{\textit{e.g.,\ }}
\newcommand{\etal}{\textit{et al.\ }}

\usepackage{amsthm}
\newtheorem{thm}{Theorem}

\usepackage{enumitem}
\setlist[itemize]{leftmargin=*,noitemsep,topsep=0pt}
\setlist[enumerate]{leftmargin=*}

\usepackage{todonotes}
\usepackage{soul}

\ifDEBUG
    \newcommand{\DA}[1]{\todo[color=white,inline]{DA:#1}}
    \newcommand{\JD}[1]{\todo[color=yellow,inline]{JD:#1}}
    \newcommand{\WJ}[1]{\todo[color=pink,inline]{WJ:#1}}
    
    \newcommand{\MC}[1]{\todo[color=cyan,inline]{MC:#1}}
    \newcommand{\GKT}[1]{\todo[color=cyan,inline]{GKT:#1}}
    
    \newcommand{\TODO}[1]{\hl{#1}}
\else
    \newcommand{\DA}[1]{}
    \newcommand{\JD}[1]{}
    \newcommand{\WJ}[1]{}
    \newcommand{\GKT}[1]{}
    
    \newcommand{\TODO}[1]{}
\fi

\usepackage{booktabs} 
\usepackage{colortbl}
\usepackage{amsmath}
\usepackage{amsfonts}
\usepackage{algpseudocode}
\usepackage{xspace}
\usepackage{multirow} 
\usepackage{balance} 
\usepackage{fancyvrb} 
\usepackage{subcaption} 
\usepackage[font=small,skip=5pt]{caption}
\usepackage{url}
\usepackage{xcolor}

\usepackage{tikz} 
\usetikzlibrary{automata} 
\usetikzlibrary{positioning} 
\usetikzlibrary{arrows} 
\usetikzlibrary{calc} 
\usetikzlibrary{patterns} 
\usetikzlibrary{snakes} 

\usepackage[group-separator={,}]{siunitx}

\VerbatimFootnotes 

\usepackage{cleveref}
\crefformat{section}{\S#2#1#3}
\crefname{figure}{Figure}{Figures}
\crefname{table}{Table}{Tables}
\crefname{listing}{Listing}{Listings}
\crefname{theorem}{Theorem}{Theorems}
\crefname{thm}{Theorem}{Theorems}
\crefname{lemma}{Lemma}{Lemmata}
\crefname{equation}{Eqt.}{Eqts.}
\crefformat{Grammar}{Grammar #1}

\usepackage{listings}

\newenvironment{RQList}{
   \setlength{\topsep}{0pt}
   \setlength{\partopsep}{0pt}
   \setlength{\parskip}{0pt}
   \begin{description}[style=unboxed]
   \setlength{\leftmargin}{1in}
   \setlength{\parsep}{0pt}
   \setlength{\parskip}{0pt}
   \setlength{\itemsep}{0pt}
   }
   {\end{description}}

\usepackage{newfloat}
\usepackage{syntax}
\usepackage{framed}

\DeclareFloatingEnvironment[
  fileext   = logr,
  listname  = {List of Grammars},
  name      = Grammar,
  placement = htp
]{Grammar}

\usepackage[ruled,vlined]{algorithm2e}

\usepackage{nicefrac}

\usepackage{graphicx}
\usepackage{caption}
\usepackage{subcaption}
\usepackage{float}
\usepackage{multirow}

\newcommand\StepSearch{Step 1}
\newcommand\StepClassifyFailure{Step 2}
\newcommand\StepClassifyAnalyzability{Step 3}
\newcommand\StepIndex{Step 4}
\newcommand\StepRAG{Step 5}
\newcommand\StepPostmortemReport{Step 6}
\newcommand\StepThematicAnalysis{Step 7}

\newcommand\TotalIncidents{2,457}
\newcommand\RAGIncidents{26}
\newcommand\NonRAGIncidents{2,431}

\usepackage{acmart-taps}
\usepackage{xcolor}
\usepackage{color}
\usepackage{rotating}
\usepackage{tabularray}
\usepackage{pbox}

\usepackage{scalerel}
\usepackage{soul}

\begin{document}



\title{FAIL: Analyzing Software Failures from the News Using LLMs}


\ifBLINDED
\author{Anonymous Author(s) \hspace{0.25cm} --- \hspace{0.25cm} Submission ID \#X}
\else
\author{Dharun Anandayuvaraj}
\orcid{0000-0001-6191-1180}
\affiliation{%
  \institution{Purdue Universityj}
  \city{West Lafayette, IN}
  \country{USA}
}
\email{dananday@purdue.edu}

\author{Matthew Campbell}
\orcid{0009-0002-3899-6117}
\affiliation{%
  \institution{Purdue University}
  \city{West Lafayette, IN}
  \country{USA}
}
\email{mcampbell@purdue.edu}

\author{Arav Tewari}
\orcid{0000-0002-1512-858X}
\affiliation{%
  \institution{Purdue University}
  \city{West Lafayette, IN}
  \country{USA}
}
\email{atewari@purdue.edu}

\author{James C. Davis}
\orcid{0000-0003-2495-686X}
\affiliation{%
  \institution{Purdue University}
  \city{West Lafayette, IN}
  \country{USA}
}
\email{davisjam@purdue.edu}
\fi

\begin{abstract}


Software failures inform engineering work, standards, regulations.
For example, the Log4J vulnerability brought government and industry attention to evaluating and securing software supply chains.
Retrospective failure analysis is thus a valuable line of software engineering research.
Accessing private engineering records is difficult, so such analyses tend to use information reported by the news media. 
However, prior works in this direction have relied on manual analysis.
That has limited the scale of their analyses.
The community lacks automated support to enable such analyses to consider a wide range of news sources and incidents.

\deleted{In this paper, we propose the \textit{\ul{F}ailure \ul{A}nalysis \ul{Investigation} with \ul{L}LMs (FAIL)} system to fill this gap.}
\added{To fill this gap, we propose the \textit{\ul{F}ailure \ul{A}nalysis \ul{Investigation} with \ul{L}LMs (FAIL)} system.}
FAIL \added{is a novel LLM-based pipeline that} collects, analyzes, and summarizes software failures as reported in the news.
FAIL groups articles that describe the same incidents.
It then analyzes incidents using existing taxonomies for postmortems, faults, and system characteristics.
To tune and evaluate FAIL, we followed the methods of prior works by manually analyzing 31 software failures.
FAIL achieved
  an F1 score of 90\% for collecting news about software failures, 
  a V-measure of 0.98 for merging articles reporting on the same incident,
  and extracted 90\% of the facts about failures.  
We then applied FAIL to a total of 137,427 news articles from 11 providers published between 2010 and 2022.
FAIL identified and analyzed \TotalIncidents{} distinct failures reported across 4,184 articles.
Our findings include:
  (1) current generation of large language
models are capable of identifying news articles that describe failures,
and analyzing them according to structured taxonomies;
  (2) high recurrences of similar failures within organizations and across organizations;
  and
  (3) severity of the consequences of software failures have increased over the past decade.
The full FAIL database is available so that researchers, engineers, and policymakers can learn from a diversity of software failures.

\TODO{
1. Verify articles in Figure 1
2. Swap citation for LLMs for Requirements Engineering
3. Fix citation in second to last sentence in 2.3 
}

\end{abstract}



\maketitle

\section{Introduction} \label{sec:introduction}

\begin{figure*}[!ht]
    \centering
    \includegraphics[width=0.90\linewidth]{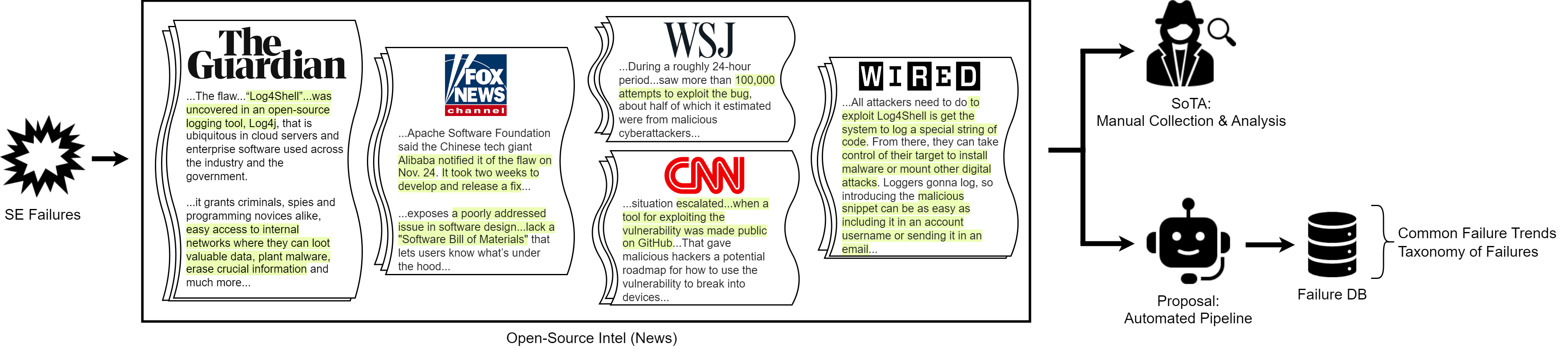}
    \caption{
    Concept of the FAIL (\ul{F}ailure \ul{A}nalysis \ul{I}nvestigation with \ul{L}LMs) project.
    Software failures are often described in news articles, \eg the examples here for the Log4j security vulnerability.
    If analysts can identify and collate related articles about failures, they can share this knowledge with software engineers and policymakers.
    Current approaches are manual. 
    We propose the FAIL system to automatically collect, group, and analyze software failures from the news.
    Our FAIL prototype collected, grouped, and analyzed these articles together for Log4j.
    \JD{Dharun, please confirm that these articles were found by FAIL and not by hand. If these are *not* done by FAIL, please update the figure to put in real articles!}
    \JD{Increase font size in figure while you're at it --- specifically the ``SE failures'' and ``SoTA'' and ``Proposal'' and so on.}
    }
    \label{fig:PaperOverview}
\end{figure*}



\JD{The first two paragraphs seem slightly too negative about prior work. Give some credit to prior SWEng work on retrospectives etc as well as to prior manual efforts to analyze failures. The second paragraph in particular should be more diligent about prior work. Basically I think we are not following the 4-paragraph ``Duality Lab Intro'' that I recommend. We could possibly do 5 paragraphs but it currently feels too detailed in the Intro. Please think about this. Perhaps move some material to \$2 or Discussion.}
Software has become pervasive and ubiquitous in modern society. 
Technologies such as the Internet of Things (IoT) and Cyber-Physical Systems have enabled software to increasingly interact with the physical world~\cite{ratasichRoadmapResilientInternet2019}.
Given the complexities of modern software systems, their diverse characteristics enable diverse failures~\cite{perrow2011software, koThirtyYearsSoftware2014}. 
Recent software failures have had catastrophic consequences~\cite{koThirtyYearsSoftware2014, wongBeMoreFamiliar2017, wongRoleSoftwareRecent2009, zollersNoMoreSoft2005, NAP11923}, and the need to reduce software failures is growing. 
As the National Academy of Engineers has emphasized~\cite{NAP11923}, software engineering failures remain commonplace and cost an estimated trillions of dollars annually~\cite{krigsmanAnnualCostIT2009}. 
We believe that one contributing factor to the high rate of software failure is the limited effectiveness and adoption of failure-aware software development practices~\cite{anandayuvarajReflectingRecurringFailures2023}.

A fundamental engineering principle is to analyze failures and act to mitigate them in the future~\cite{petroskiDesignParadigmsCase1994, dalcherFallingPartGrowing1994}.
This principle has been successful in many engineering disciplines, and has contributed to the low failure rates of, e.g., buildings, aircraft, and nuclear reactors. \GKT{I'm worried about the contrast statement below. After all, many of these have software. Aircraft and space machines (e.g. space shuttle, remember that?) have a long history of software and even various mitigations. It may be important to have a working definition/assumption of how SE process plays out in different domains. A Beoing 787 is more important than Microsoft Word (I think).}
In contrast, this principle is rarely followed consistently as part of the software engineering process \textit{within} organizations (intra-organizational learning)~\cite{vieiraTechnicalManagerialDifficulties2019, kasiPostMortemParadox2008, dingsoyrPostmortemReviewsPurpose2005, glassProjectRetrospectivesWhy2002} or \textit{across} organizations (inter-organizational learning)~\cite{johnsonSoftwareSupportIncident2000, normanCommentaryHumanError1990}.
\GKT{Conjecture: Organizations that have certification requirements may be better at embracing (boring) SE processes. They are not immune to failure as we learned with Boeing's mishaps but their processes lend themselves to discovery and isolation more readily. Telecommunications equipment is one that I experienced. They got more excited about process after undergoing ISO 9000 certification...}
Although organizations may be unwilling to publicly disclose their own failures, news articles and other kinds of grey literature could provide sufficient information on failures to facilitate inter-organizational learning. \GKT{Not necessarily in the scope of this paper, but it would be nice to know what happens after "outed" in the news, especially in public companies with accountability to shareholders. Do they take action, hold themselves accountable, and report to the stakeholders what they plan to do to improve their processes?}

News agencies often report on public-facing engineering failures: negatively, emphasizing the problem (defects), or positively, highlighting the resolutions  (mitigations)~\cite{astvanshEffectsNewsMedia2022}.
Specifically, news articles reporting on software failures often contain information related to system and design level causes, impacts, and lessons learned from an incident~\cite{anandayuvarajReflectingRecurringFailures2023,rahmanIdentificationSourcesFailures2009, koThirtyYearsSoftware2014, raharjanaUserStoryExtraction2019}.
Additionally, software incidents often result from a combination of organizational, managerial, technical, sociological, and political factors, and preventing such incidents necessitates addressing all underlying causes, extending beyond code-level issues~\cite{leveson1995safeware, reason2016managing}.
Suitably, news articles reporting on software failures often contain such contextual information.
For example, the Cyber Safety Review Board (CSRB)~\cite{CyberSafetyReview} in the Cybersecurity and Infrastructure Security Agency (CISA)~\cite{HomePageCISA} published a failure report about the Log4j incident~\cite{CSRBReportLog4J}, in which they note news articles that reported failure information surrounding the incident. 
As illustrated in~\cref{fig:PaperOverview}, the report lists articles containing failure information from The Guardian~\cite{associatedpressRecentlyUncoveredSoftware2021}, Fox News~\cite{associatedpressChineseIranianHackers2021}, The Wall Street Journal~\cite{mcmillanSoftwareFlawSparks2021}, CNN~\cite{lyngaasDHSWarnsCritical2021}, and WIRED~\cite{newmanInternetFire}.

\GKT{This transition needs to be more smooth opening sentence. Content of paragraph is effective otherwise. Are we saying that the fact that these news sources are open that it amounts to an "open" data source of sorts? In any event, I'd like a working definition and how it connects to the cited paper. In general, we don't want to make people read other papers to establish an understanding of an undefined term.}

\added{Failures reported by the news often focus on critical failures that have a substantial effect on society.}
Such news data comprise ``Open-Source Intelligence''~\cite{therecordedfutureteamWhatOpenSource2022}, and may be used by organizations, government bodies, and academics to formulate best practices, draft regulations, or discover research directions~\cite{gillWhatOpenSourceIntelligence2023}.
In the past, high-profile software incidents such as Therac25~\cite{turner1993investigation}, Stuxnet~\cite{farwellStuxnetFutureCyber2011}, and Boeing 737 MAX crashes~\cite{johnston2019boeing}, have informed software engineering practices, guidelines, and policymaking~\cite{turner1993investigation,farwellStuxnetFutureCyber2011,johnston2019boeing}.
\deleted{Researchers have theorized that such news may discipline organizations, leading them to make socially responsible decisions in the context of safety~\cite{astvanshEffectsNewsMedia2022}.}
\added{Recently, high-profile incidents such as Crowdstrike~\cite{WidespreadITOutage2024}, Colonial Pipeline~\cite{AttackColonialPipeline2023}, Log4j~\cite{ApacheLog4jVulnerability2022}, and SolarWinds~\cite{RemediatingNetworksAffected2021}, continue to shape software engineering practices, guidelines, and policymaking.
These failures are often covered in detail by the media precisely because they are so consequential. 
These failures impact society in substantial ways (\eg physical harm, property loss, financial loss, etc.).
These failures are important, and the software engineering community should understand and address them.}
Such failures could be used to influence rationales for software design decisions, and enable failure-aware software development practices~\cite{anandayuvarajIncorporatingFailureKnowledge2023}. 

However, current approaches to obtaining open source intelligence \GKT{This is why you need a definition, because the \eg is not a definition here and feels disconnected.} (\eg studying news articles reporting on failures) require costly expert manual analysis which limits the scale of analysis.
For example, the only prior large-scale study related to software failures relied on manual analysis to 
study the consequences of 3,977 software problems from the news~\cite{koThirtyYearsSoftware2014}, which reportedly took 1,000 person-hours over 2 months.
A similar work manually 
analyzes 347 communication and IT infrastructure failures from the RISKS digest~\cite{rahmanIdentificationSourcesFailures2009}. 
\GKT{Was any sort of dataset made from this study? I haven't read ahead yet, but is FAIL able to identify the same failures and, if so, is it in dramatically less time? That result would be nice.}
There is a gap in current software engineering research to efficiently gather, analyze, and report insights from open-source information on recent software failures to enable failure-aware software development practices.

To reduce the costs and improve the scalability of manual analysis, we propose an automated pipeline that employs Large Language Models (LLMs) to collect, analyze, and report insights from software failures reported in the news (\cref{fig:PaperOverview}).
Using this pipeline we conducted a large-scale systematic study of recent software failures as reported in the news. 
\ul{Our main contributions are}:

\begin{enumerate}    
    \item We design and implement FAIL, an automated approach for large-scale analysis of software failures from news using LLMs.
    \item We publish a database of postmortems describing software failures from 2010-2022, which can be updated at the cost of approximately \$50 per year of data at present LLM costs.
    \item We provide a large-scale report on the common sources, impacts, repair recommendations, etc. of these software failures. 
\end{enumerate}

\textbf{Significance: } 
\added{We design, implement, and evaluate FAIL,
a novel LLM-based pipeline that collects, analyzes, and reports insights from software failures reported in the news.}
\deleted{Our work shows the effectiveness of an automated approach in enabling large-scale news failure analysis.}
FAIL is fully automated so it can become an ongoing community resource with regular updates.
The resulting database can inform software engineering practice and research, policy making, and education.

\section{Background and Related Work} \label{sec:background}

Here we discuss
  failures as a feedback mechanism (\cref{sec:Background-FailuresAsFeedback}),
  with
  news articles as a source (\cref{sec:Background-News}),
  and
  large language models as a tool (\cref{sec:Background-NLP}).

\subsection{Failures as Feedback} \label{sec:Background-FailuresAsFeedback}
\subsubsection{Definition of software failures}
Engineers expect some \ul{defects}~\cite{kuutilaTimePressureSoftware2020, costello1984software}, but try to eliminate severe defects that may cause \ul{incidents}: undesired, unplanned, software-induced events that cause substantial loss~\cite{leveson1995safeware}.
Whether severe defects are caught internally or as incidents, their presence is a \ul{failure} indicating flawed software engineering process.
Not all defects are the same, and understanding failures assists with failure mode analysis~\cite{reiferSoftwareFailureModes1979,song2012applying,ishimatsu2010modeling} and helps researchers and engineers to identify and prioritize risks~\cite{fairbanksJustEnoughSoftware2010}.

\subsubsection{The study of failures in SE}
Studying engineering failures enables successful design~\cite{petroskiDesignParadigmsCase1994}.
Software researchers have extensively studied 
\deleted{failures}
\added{the presence of defects~\cite{luLearningMistakesComprehensive2008, gunawi2014bugs, makhshariIoTBugsDevelopment2021} and 
vulnerabilities (defects that can be exploited by attackers to compromise a system's security) ~\cite{holzingerInDepthStudyMore2016, jimenezEmpiricalAnalysisVulnerabilities2016a, shuStudySecurityVulnerabilities2017}} in open-source software.
\added{These studies analyze defects and vulnerabilities of open-source software in isolation and trace the defect to flaw(s) in the implementation, although not always considering deployment context or severity of impact~\cite{amusuo2022SoftwareFailureAnalysis}.}
However, we believe the most interesting failures occur in contexts where engineers followed robust engineering processes~\cite{kalu2023reflecting}.
Thanks to standards (\eg IEC 61508~\cite{IEC6150812010}) and regulations (\eg the EU GDPR~\cite{gdpr2018}, the US FDA FD\&C Act section 524B~\cite{FDA2023Cybersecurity}), those contexts are generally commercial.

Only a few works have studied commercial software failures in detail.
\GKT{I may have answered my own question. Still, the setup leading to this paragraphs strongly suggests this as well. Should we just come out and say that there are few works that have studied it and only cite those? I think the defects citations in the previous belong in the first subsection, which I would relable to include the word defects, since you talk about defects there.}
Commercial failure information is difficult to obtain, so the normal approach is to identify relevant news articles (\eg via searches) and then analyze them by hand.
For an early example, Wallace \etal studied 342 software failures in medical devices from the FDA medical device failures database from 1983 to 1997~\cite{wallaceLessons342Medical1999}.
Researchers commonly study
  news articles~\cite{koThirtyYearsSoftware2014,wongBeMoreFamiliar2017,anandayuvarajReflectingRecurringFailures2023,howardAnalysisSecurityIncidents1997, bertlNewsAnalysisDetection2019},
  and have also analyzed data from
  the RISKS digest~\cite{neumannComputerRelatedDisastersOther1986}, 
  and public postmortems~\cite{sillitoFailuresFixesStudy2020a}.
Many of these works consider dozens or hundreds of failures, while others present failure case studies to inform software engineering education, practice, and policymaking~\cite{leveson1995safeware, leveson2016engineering}.
Although these failure analysis studies have advanced the software engineering field~\cite{millett2007software, leveson1995safeware}, they require costly expert manual analysis which limits their scale and update rate. 
Studies have not shown how to \ul{automatically} gather, analyze, and report insights from news articles about software failures. 

\subsubsection{Databases of failures for inter-organizational learning}
Other engineering fields benefit from periodically-updated, large-scale, industry-wide databases providing postmortem information about failures.
For example, there are databases of failures in
  medical devices~\cite{commissionerMedWatchFDASafety2022}, 
  aviation~\cite{ASRSAviationSafety},
  aerospace~\cite{NASA2023LessonsLearned}, 
  railways~\cite{RailAccidentInvestigation2017},
  and
  chemicals~\cite{InvestigationsCSB}.
These enable practitioners in those fields to learn from past failures across organizations.
Within software engineering, such databases are limited to
  vulnerabilities (\eg CVEs ~\cite{CVECommonVulnerabilities})
  and
  open-source defects (\eg BugSwarm~\cite{BugSwarm}).
There is also the RISKS Digest, a manually maintained list of computing related incidents and risks~\cite{neumannRISKSDigest}.
Although researchers have highlighted the need for large, up-to-date databases of software failures~\cite{normanCommentaryHumanError1990,johnsonSoftwareSupportIncident2000}, none exists, perhaps due to technology gaps in scaling up manual analysis~\cite{johnsonSoftwareSupportIncident2000}.

\subsection{News Articles as a Data Source} \label{sec:Background-News}

\subsubsection{Definition and Biases}
The notion of a news source can be broad, including any ``open-source intelligence''~\cite{therecordedfutureteamWhatOpenSource2022,gillWhatOpenSourceIntelligence2023}.
Examples include
  the mainstream news media (\eg The New York Times, CNN),
  public information from companies (\eg public postmortems and vulnerability writeups) and other organizations (\eg wikis and databases in open-source repositories),
  and
  individual views expressed in blogs and podcasts.
A news article typically summarizes and reports on events rather than being the primary source of the information.
News articles provide historic and real-time information, presenting an opportunity for researchers to explore and analyze dynamic events and trends across diverse domains.
There is a large volume of articles each day, some of which contain insight into industry failures and software in particular.

We acknowledge the limitations of news articles, notably source bias~\cite{manning2000news} and editorial bias~\cite{rodrigo-ginesSystematicReviewMedia2024}.
For example, news articles include information from multiple sources, including
  official statements (\ie biased organizational perspectives),
  government reports and investigations (\ie delayed, independent analyses),
  expert opinions and analyses (\ie independent analysis varying in objectivity and accuracy),
  internal leaks and whistle-blowers (\ie internal organizational views of unclear reliability),
  and user reports or social media (\ie public perspectives lacking technical insight and of unclear reliability)~\cite{manning2000news}. 
The author of the news article, \eg a journalist, may introduce their own biases, notably
 selection bias (preferring sensational incidents),
 framing bias (adding narration for audience engagement),
 and
 information gaps (authors may provide incomplete or inaccurate analysis)~\cite{rodrigo-ginesSystematicReviewMedia2024}. 
 

\subsubsection{Use of News in Other Fields.}
News is used to study failures and inform practices in various fields. 
News reports have been used as primary data sources in works to identify cascading effects of infrastructure failures on other infrastructures and stakeholders~\cite{zhouDelineatingInfrastructureFailure2020}, and 
to identify risks and conduct a quantitative analysis to study cost overrun in rail transit projects~\cite{gaoNaturalLanguageProcessing2022}.   
In traffic engineering, news has been used to identify traffic incidents and notify citizens~\cite{riveraNewsClassificationIdentifying2020}, and to study the characteristics and reasons for first responder-involved incidents~\cite{yangAnalysisFirstResponderinvolved2023}.
In environmental engineering, news has been used to study chemical pollution~\cite{combyHowChemicalPollution2014}.
In public health, news has been used to study food safety~\cite{qiangApplicationContentAnalysis2011}.
In civil engineering field, the Engineering News Record gives examples of failures weekly~\cite{ENRUsEngineering}, and is used to form case studies to inform education and practices~\cite{delatteForensicsCaseStudies2002a}. 
These works show that incident information (\eg causes, impacts, stakeholders) can be extracted and studied from the news.
However, at the moment they rely on manual analysis, limiting their scale and update frequency.

\subsection{NLP and LLMs for Software Engineering} \label{sec:Background-NLP}

Natural Language Processing (NLP) has been leveraged for various phases of the Software Development Life-Cycle (SDLC).  
Classical NLP tools have been applied in
  specification (requirements extraction)~\cite{zhaoNaturalLanguageProcessing2022};
  design (system modeling)~\cite{bajwaObjectOrientedSoftware};
  implementation (code generation)~\cite{ernstNaturalLanguageProgramming2017};
  testing (identifying risks)~\cite{garousiNLPassistedSoftwareTesting2020,vijayakumarAutomatedRiskIdentification2017};
  and
  maintenance (analyzing user feedback)~\cite{panichellaHowCanImprove2015}.
  
In this paper, we apply a recent innovation in NLP --- Large Language Models (LLMs) --- to study software failures.
LLMs are neural network-based language models that ``understand'' and generate text~\cite{brantsLargeLanguageModels}.
LLMs have been applied to software engineering tasks~\cite{houLargeLanguageModels2023, fanLargeLanguageModels2023},
\eg
  requirements~\cite{rahmanAutomatedUserStory2024}, 
  design~\cite{whiteChatGPTPromptPatterns2023c}, 
  implementation~\cite{jiangSelfplanningCodeGeneration2023}, 
  testing~\cite{khanfirEfficientMutationTesting2023}, 
  deployment~\cite{shypulaLearningPerformanceImprovingCode2023}, 
  and
  maintenance~\cite{xiaKeepConversationGoing2023}. 

The closest lines of work to ours have applied NLP and LLMs for incident management.
Studies using NLP techniques have
  mined dialogues of IT staff~\cite{azadPickingPearlSeabed2022},
  found entities in incident descriptions~\cite{shettyNeuralKnowledgeExtraction2021a}, 
  pulled root causes from incident reports~\cite{saha2022mining},
  and categorized IT incident tickets~\cite{silvaMachineLearningIncident2018}.
More recently, LLMs have been applied to incident management.
These works have used cloud incident data, often based on internal (not public) reports. 
On that class of data, LLMs can
  assess the impact scope and summarize cloud outages~\cite{jin2023assess}, 
  identify root causes and mitigation steps~\cite{ahmedRecommendingRootCauseMitigation2023a,chenAutomaticRootCause2024, zhangAutomatedRootCausing2024},
  and
  recommend further queries~\cite{jiang2024xpert}.
Our work addresses similar tasks but is focused on publicly available data in the news, more limited in detail but more accessible across organizations. 

\section{FAIL: Design and Implementation} \label{sec:DesignImplement}

\JD{@DHARUN I did big changes in this section, PTAL for correctness etc}

\JD{Dharun, make sure all missing references are in. I caught a bunch but I see that RAG is still missing.}

Let us summarize the background and related work.
Engineering failures are often covered in the news.
News articles have known biases, but are sometimes the only source of insight into failures of high societal impact.
Thus, researchers in many engineering disciplines --- including software engineering --- have extracted knowledge from this data source.
Current approaches to this are largely manual.
Meanwhile, large language models (LLMs) have shown promise in automatically analyzing natural-language descriptions of failures, notably postmortems written by engineers.

LLMs have not been applied to software failures in the news.
\added{However, to use LLMs for large-scale failure analysis, we must measure their ability to analyze failures in the news, and assess their degree of inaccuracy (hallucination).}
To this end, we propose an automated pipeline, \textbf{FAIL}, that uses LLMs to collect and analyze software failures in the news.
FAIL is conceptually depicted in~\cref{fig:PaperOverview}.
We address three requirements:

\begin{enumerate}
\item \textit{R1-Scale:} FAIL should handle years' worth of news data.
\item \textit{R2-Accuracy:} FAIL should be able to correctly
  (a) identify news articles about software failures,
  (b) merge related articles,
  and
  (c) analyze them for postmortem data.
\item \textit{R3-Cost:} FAIL should do so cheaply, \eg $<$\$1 per incident, orders of magnitude less than the cost of manual approaches.
\end{enumerate}

\begin{figure*}[!ht]
    \centering
    \includegraphics[width=0.80\linewidth]{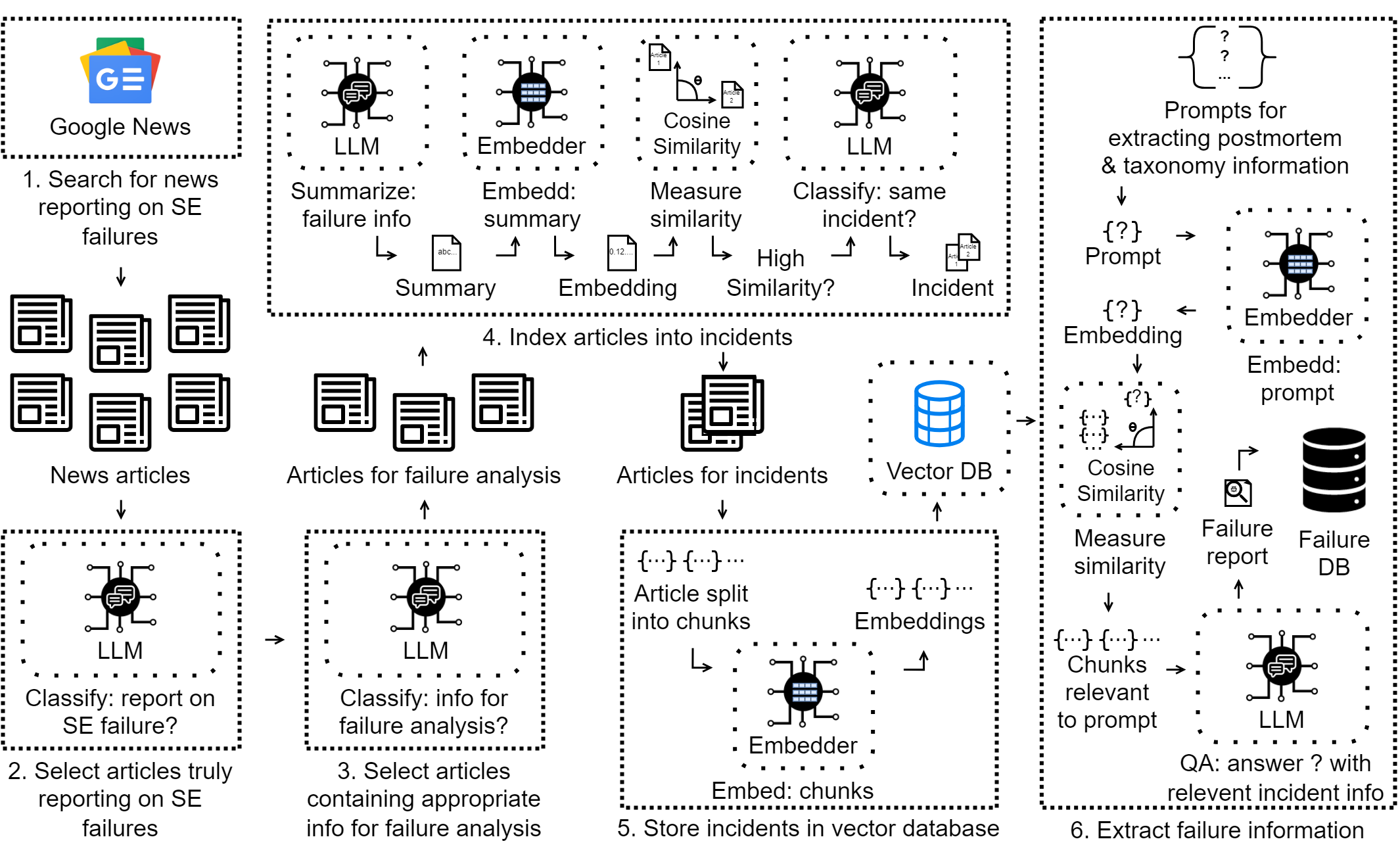}
    \caption{
    Overview of the proposed FAIL pipeline to collect, analyze, and report insights from software failures reported in the news.
    FAIL has 6 components.
    In Step 1, we search for news based on criteria such as keywords, sources, and timeframe.
   In Step 2, we remove articles that are not about software failures.
   In Step 3, we remove articles that do not have enough information to analyze.
   In Step 4, we merge articles reporting on the same incident.
   In Step 5, we use RAG techniques to handle long incidents.
   In Step 6, we create a failure report for each incident.
    \JD{Dharun, in the figure, write ``STEP X'' in each place, and in bold and perhaps slightly larger font.}
    }
    \vspace{-0.30cm}
    \label{fig:MethodOverview}
\end{figure*}

\cref{fig:MethodOverview} shows our FAIL design.
Components are detailed next.

\subsection{Approach for each FAIL Component}  \label{Pipeline: Description}

\textbf{\StepSearch{}---Initial search:}
We began by creating an initial criteria for news sources and keywords to search their databases.
%
%
Following prior work (\cref{sec:Background-News}), we used globally popular news sources~\cite{majidTop50Biggest2023} as well as common sources used in the RISKS Digest~\cite{rahmanIdentificationSourcesFailures2009}.
The resulting 11 \ul{news sources} were: Wired, The New York Times, CNN, Daily Mail, The Guardian, BBC News, Fox News, The Washington Post, CNET, Reuters, and AP News. 
We compiled 13 \ul{keywords} related to software failure~\cite{TestingExperienceTe2008}, such as ``software flaw'', ``software bug'', and ``software crash''.
We used Google News as the search engine.
We searched for news from these sources, matching these terms, from 2010 to 2022. 
For each search result, we used a news scraper~\cite{ou-yangNewspaper3kArticleScraping2023} to scrape the text content of each article. 

\textbf{\StepClassifyFailure{}--Relevance:}
The search results contain news articles that may be irrelevant. 
To filter out such articles, FAIL prompts an LLM for whether the article actually reports on software failures.

\textbf{\StepClassifyAnalyzability{}--Level of detail:}
The relevant articles may not have enough detail to conduct failure analysis, which could lead to low performance during failure analysis~\cite{singla2023empirical}.
To filter out such articles, FAIL prompts an LLM for whether the articles contain enough information for failure analysis.

\textbf{\StepIndex{}--Merge:}
A high-profile failure (\eg the Boeing 737 MAX crashes~\cite{gellesBoeing737Max2019, luoInternetThingsLoT2020}) may be covered by many articles.
The articles may have different information, \eg based on different sources, so we want to analyze them together for a full understanding of the failure.
FAIL merges these articles into a single \textit{incident record}, similar to prior work~\cite{koThirtyYearsSoftware2014} but applying LLM technology rather than classical NLP.
To reduce costs, our approach proceeds in two stages: approximation and a follow-up check.
For the approximation, FAIL prompts the LLM to create a summary of the failure as reported in each article, and then uses OpenAI's sentence transformer (text-embedding-ada-002) to convert the summary into an embedding.
FAIL then calculates the cosine similarity of the summary embedding against that of existing incidents in the database.
If the similarity exceeds a threshold, we do a follow-up check with the LLM.
That check uses the summary of the article and the existing incident from the DB.
If the article is deemed similar, we associated it with the existing incident, else we create a new incident.

\definecolor{MineShaft}{rgb}{0.2,0.2,0.2}
{
\begin{table}
\caption{
Failure information per incident,
aligned with taxonomies from prior work on software failure analysis.
We introduced an IoT-specific taxonomy as many articles discussed IoT systems.
}
\label{table:QueryCatalogue}
\centering
\small
\begin{tblr}{
  width = \linewidth,
  colspec = {Q[88]Q[854]},
  cell{1}{1} = {r=9}{c},
  cell{10}{1} = {r=10}{c},
  cell{20}{1} = {r=3}{c},
  cell{23}{1} = {r=2}{c},
  vlines,
  hline{1,10,20,23,25} = {-}{},
  hline{2-9,10-19,21-22,24} = {2}{},
}
\begin{sideways}\textit{Postmortem}~\cite{dingsoyrPostmortemReviewsPurpose2005}\end{sideways}                
& Timeline                                                           \\
& System
        \\
& Responsible Organization
        \\
& Impacted Organization
        \\
& Software causes                                                    \\
& Non-software causes                                                \\
& Impacts                                                           \\
& Prevention \& Fixes                                             \\
& References
        \\
\begin{sideways}\textit{Software: Taxonomy of Faults}~\cite{avizienisBasicConceptsTaxonomy2004a}\end{sideways} 
& Recurring: At one organization, or multiple \\
& Phase: Design, Operation                                                     \\
& Boundary: Internal, External                                        \\
& Nature: Human, non-human actions                                \\
& Dimension: Hardware, Software                                       \\
& Objective: Malicious, Non-malicious                                 \\
& Intent: Poor decisions, Accidental decisions                                  \\
& Capability: Developmental incompetence, Accidental                                \\
& Duration: Permanent, Intermittent                         \\
& Behaviour: Crash, Omission, Timing, Value, Byzantine \\
\begin{sideways}{\pbox{1.5cm}{\textit{IoT System:} \newline \textit{Layer}~\cite{meloPathologyFailuresIoT2021}}}\end{sideways}     
& Perception: Sensors, Actuators, Processing, Network, Embedded software \\
& Communication: Link, Connectivity                                   \\
& Application            \\        
\begin{sideways}\textit{Other}~\cite{koThirtyYearsSoftware2014}\end{sideways} 
&
Consequence:
Death, Harm, Property, Delay, Basic, None
\\
& Domain (12 total):
IT, Transport, Manufacturing, etc.
\\
\end{tblr}
\end{table}
}


\textbf{\StepRAG{}--Handling long articles (RAG):}
LLMs have a context window (\eg $\sim$12K words for the LLM we used).
Because we merge articles into incident records for analysis, our data on some incidents may exceed the context window. 
In our database, \RAGIncidents{} incidents (out of \TotalIncidents{}) exceeded this context window. 
To accommodate such incidents, FAIL uses Retrieval-Augmented Generation (RAG).
For each article, FAIL stores embeddings in a vector database. 
A vector database supports similarity searches using indexing techniques and similarity metrics, allowing FAIL to reference multiple articles within a context window.
Our approach is typical of RAG:
  we chunk the article into paragraphs,
  label them with incident id, article id, and their order in the article,
  and embed them using OpenAI's sentence transformer. 
Based on similarity to each prompt in Step 6, the most relevant chunks from each incident are provided as context to the prompt.

\textbf{\StepPostmortemReport{}--Postmortem analysis:}
Finally, FAIL analyzes each incident.
We developed prompts to extract information prescribed by several prior works on failure analysis. 
The fields are shown in~\cref{table:QueryCatalogue}.
FAIL collects the articles for each incident as context, and prompts the LLM for each field of the failure report.

The prompts were engineered to extract information relevant to postmortems~\cite{dingsoyrPostmortemReviewsPurpose2005}, a taxonomy of software faults~\cite{avizienisBasicConceptsTaxonomy2004a}, and additional details about the incident~\cite{meloPathologyFailuresIoT2021, koThirtyYearsSoftware2014}.
The failure report contains open-ended (postmortem) fields and multiple-choice (taxonomy) fields.
For the open-ended fields, the prompts contained instructions to \textit{extract} information from the incidents relevant to the fields and their definitions. 
For extracting the timeline of incidents, further instructions (including chain-of-thought prompting~\cite{wei2022chain} and few-shot prompting~\cite{brown2020language}) were added to guide the LLM into estimating the timeline of incidents.
For the multiple-choice fields, we likewise applied chain-of-thought prompting.
For each field, we first prompt the LLM to extract information from the incidents related to all of the options,
and
then we prompt the LLM to mark whether each option is supported by the information it extracted. 

To illustrate, we show the prompt to extract information about the \textit{system that failed}.
This prompt is for row 2 in~\cref{table:QueryCatalogue}.

\begin{tcolorbox} [width=\linewidth, colback=green!30!white, top=1pt, bottom=1pt, left=2pt, right=2pt]
What system(s) failed in the software failure incident?
If specific components or models or versions failed include them. Return the products, systems, components, models and versions that failed in a numbered list (with citations in the format: [\#, \#, ...]).
\end{tcolorbox}

\subsection{Component Development and Validation} \label{sec:Component Development and Validation}

To develop and validate each component, we developed a ground-truth dataset by manually performing each step of the pipeline.
\added{Two analysts independently analyzed randomly selected news articles until acceptable inter-rater agreement was reached in each phase of the analysis pipeline.
When they disagreed, they resolved through discussion, sometimes revising the definitions (\eg in the taxonomies we used).
The artifact contains a record of these discussions and resulting changes.}
Our human analysts were qualified: they were students in computing (1 graduate, 1 undergraduate) and both have native-level English.
Labeling the ground-truth dataset took 4 weeks with 2 analysts, or $\sim$320 person-hours.  

We describe some details for each Step.
Since FAIL uses learned components, we do not expect perfect performance.
Our goal is to understand the degree of uncertainty present in our subsequent analysis of the FAIL DB (\cref{sec:FailureAnalysis}).

\textbf{Steps 2-4: Relevance, detail, clustering:}
The underlying data are somewhat sparse: the Google news queries return 137K articles, of which our final pipeline found 6K to be relevant and 4K to be analyzable.
Thus, had we used a truly random sample from Google News we would have had to examine many irrelevant articles.
Instead, for these steps, we ran the FAIL pipeline and used stratified sampling based on FAIL's opinion in order to get articles for our manual ground truth.
We used stratified sampling as follows:
  (1) we collected articles from 30 incidents (the articles that went through \StepIndex{} of the pipeline and were indexed into incidents),
  and
  (2) we collected 30 articles which were (2.a) classified as \textbf{not} reporting on a software failure - articles that went through \StepClassifyFailure{} of the pipeline, and (2.b) classified as reporting on a software failure AND classified as \textbf{not} containing sufficient failure information to conduct failure analysis - articles that went through \StepClassifyAnalyzability{} of the pipeline. 

For Steps 2-3, we did this twice.
\textit{First} we focused on 2010-2016.
This sample had 77 articles.
The analysts labeled these for relevance and level of detail and we iteratively improved the prompts accordingly.
For example, we removed the phrase ``failure analysis'' from these prompts and added more specific criteria instead.
These data were then discarded to avoid overfitting.

\textit{Second}, we randomly sampled 76 articles from 2010-2022.
On this set, the analysts had internal 92\% agreement for Step 2, and 80\% agreement for Step 3.
After resolving disagreement, 50 of the 76 articles were related to software failures, and 45 as having enough information.
On this sample, the FAIL prompts achieved F$_1$ scores of 90\% and 91\%, respectively.
Confusion matrices are given in~\cref{tab:step2step3Confusion}.

{
\begin{table}[htbp]
    \centering
    \caption{
    Confusion matrices for Step 2 and Step 3.
    }
    \small
    \label{tab:step2step3Confusion}
    \begin{subtable}[c]{0.48\linewidth}
        \centering
        \caption{Step 2: About a software failure?}
        \begin{tabular}{cccc}
             &  & \multicolumn{2}{c}{\textit{Predicted}} \\
            \multirow{4}{*}{\begin{sideways}\textit{Manual}\end{sideways}} &  & TRUE & FALSE \\
            \cline{3-4}
             & TRUE & 50 & 0 \\
             & FALSE & 11 & 15 \\
            \cline{3-4}
        \end{tabular}
        
    \end{subtable}%
    \hfill
    \begin{subtable}[c]{0.45\linewidth}
        \centering
        \caption{Step 3: Analyzable?}

        \begin{tabular}{cccc}
             &  & \multicolumn{2}{c}{\textit{Predicted}} \\
            \multirow{4}{*}{\begin{sideways}\textit{Manual}\end{sideways}} &  & TRUE & FALSE \\
            \cline{3-4}
             & TRUE & 39 & 6 \\
             & FALSE & 1 & 4 \\
            \cline{3-4}
        \end{tabular}
        
    \end{subtable}
\end{table}
}

For \textit{Step 4}, we evaluated on our manual analysis as well as to clustered articles from prior work~\cite{koThirtyYearsSoftware2014}.
We use common metrics for evaluating clustering~\cite{rosenberg2007v}:
  \textit{homogeneity} measures whether each incident contains only articles that belong to a single incident,
  \textit{completeness} for measuring whether all articles that belong to the same incident are indexed into the same incident,
  and
  \textit{V-measure} for measuring the balance between homogeneity and completeness.
Results are in~\cref{tab:Step4Performance}.
The performance of our initial prompts (prior to formal evaluation) was sufficiently good that we did not refine the Step 4 prompts after this.

{
\begin{table}[htbp]
    \centering
    \caption{Performance of FAIL \StepIndex{}: merging articles into incidents.}
    \label{tab:Step4Performance}
    \small
    \begin{subtable}[c]{0.45\linewidth}
        \centering
        \caption{On the 30 incidents (45) articles from our own pipeline.}
        \label{table: IndexingOurData}
        \begin{tabular}{lc} 
        \toprule
       \textbf{Metric}   & \textbf{Value} \\
       \toprule
            Homogeneity & 0.9686 \\
            Completeness & 0.9999 \\
            V-measure & 0.9841 \\
            \bottomrule
        \end{tabular}
    \end{subtable}
    \hfill
    \begin{subtable}[c]{0.45\linewidth}
        \centering
        \caption{On the 81 incidents (536) articles compiled from Ko \etal~\cite{koThirtyYearsSoftware2014}}
        \label{table: IndexingKoData}
        \begin{tabular}{lc}
        \toprule
       \textbf{Metric}   & \textbf{Value} \\
       \toprule
            Homogeneity & 0.9780 \\
            Completeness & 0.9292 \\
            V-measure & 0.9530 \\
        \bottomrule
        \end{tabular}
    \end{subtable}
\end{table}
}

\textbf{Step 6: Failure analysis}
For the multiple-choice (taxonomy) fields, the human analysts independently selected options based on the definitions for each field. 
We conducted 6 iterations of taxonomization until we reached definitions that resulted in a percent agreement of 80.65\%, as reported in~\cref{table: TaxonomyAgreement}.
Changes include: 
  extending the definitions with detailed explanation,
  and
  shifting the focus of the definitions from fault to failure.
We report the final percent agreement between the two analysts for each of the multiple-choice fields in~\cref{table: PostmortemReportEvaluation}. 

{
\begin{table}[htbp]
\centering
\caption{
\% agreement between the analysts taxonomizing incidents.
}
\label{table: TaxonomyAgreement}
\small
\setlength{\tabcolsep}{2pt} 
\noindent 
\begin{center} 
\begin{tabular}{@{}|l|c|c|c|c|c|c|c|@{}} 

Iteration            & 1    & 2   & 3   & 4   & 5   & 6   & Final \\
\midrule
Incidents  & 5    & 3   & 3   & 3   & 3   & 3   & 31   \\
Agreement    & 49.09\% & 43.59\% & 58.97\% & 53.85\% & 46.15\% & 89.74\% & 80.65\% \\

\end{tabular}
\end{center}
\vspace{-0.3cm}
\end{table}
}

The analysts manually conducted \StepPostmortemReport{} to analyze and create failure reports for the 30 incidents described earlier.
For the open-ended (postmortem) fields, the manual analysts independently extracted information from incidents. 
The two analysts discussed disagreements until agreement was reached. 
For this manual analysis, we report the inter-rater agreement with the proportion of overlapping facts between the two analysts and the proportion of discarded facts with respect to the number of agreed-upon facts reached by consensus in~\cref{table: PostmortemReportEvaluation}.

\begin{table*}
\caption{
  Performance of pipeline for \StepPostmortemReport{} to analyze incidents.
  The inter-rater agreement for open-ended (postmortem) fields are reported as the ratio of the overlapping facts between the two analysts to facts reached by agreement.
  } 
\label{table: PostmortemReportEvaluation}
\resizebox{\columnwidth*2}{!}{%
\begin{tabular}{|l|l|l|l|l|l|l|l|}
\hline
\textbf{Field} & \textbf{\# Facts} & \textbf{Inter-rater Agreement} & \textbf{LLM Overlapping} & \textbf{LLM Missing} & \textbf{LLM Add \& Relevant} & \textbf{LLM Add \& Irrelevant} & \textbf{LLM Incorrect} \\ \hline
Time                   & 27              & 100\%        & 93\%         & 4\%         & 37\%                     & 0\%                        & 4\%           \\ \hline
System                 & 41              & 71\%         & 100\%        & 0\%         & 73\%                     & 0\%                        & 0\%           \\ \hline
Responsible Org        & 43              & 63\%         & 88\%         & 12\%        & 9\%                      & 2\%                        & 0\%           \\ \hline
Impacted Org           & 46              & 80\%         & 96\%         & 4\%         & 17\%                     & 0\%                        & 0\%           \\ \hline
SE Causes               & 55              & 58\%         & 89\%         & 11\%        & 16\%                     & 2\%                        & 0\%           \\ \hline
N-SE Causes              & 47              & 40\%         & 68\%         & 32\%        & 45\%                     & 13\%                       & 0\%           \\ \hline
Impacts                & 93              & 54\%         & 90\%         & 9\%         & 43\%                     & 4\%                        & 1\%           \\ \hline
Preventions \& Fixes   & 70              & 49\%         & 83\%         & 17\%        & 90\%                     & 29\%                       & 0\%           \\ \hline
References             & 117             & 49\%         & 88\%         & 12\%        & 39\%                     & 0\%                        & 0\%           \\ \hline
Recurring              & 26              & 81\%         & 100\%        & 0\%         & 31\%                     & 4\%                        & 4\%           \\ \hline
Phase                  & 52              & 81\%         & 98\%         & 2\%         & 8\%                      & 6\%                        & 0\%           \\ \hline
Boundary               & 52              & 77\%         & 98\%         & 2\%         & 8\%                      & 0\%                        & 0\%           \\ \hline
Nature                 & 46              & 87\%         & 98\%         & 2\%         & 24\%                     & 2\%                        & 0\%           \\ \hline
Dimension              & 43              & 94\%         & 100\%        & 0\%         & 5\%                      & 0\%                        & 0\%           \\ \hline
Objective              & 44              & 94\%         & 77\%         & 23\%        & 5\%                      & 0\%                        & 0\%           \\ \hline
Intent                 & 30              & 71\%         & 80\%         & 20\%        & 23\%                     & 10\%                       & 0\%           \\ \hline
Capability             & 32              & 65\%         & 88\%         & 13\%        & 41\%                     & 16\%                       & 0\%           \\ \hline
Duration               & 31              & 87\%         & 100\%        & 0\%         & 13\%                     & 3\%                        & 0\%           \\ \hline
Behaviour              & 32              & 84\%         & 81\%         & 19\%        & 63\%                     & 41\%                       & 3\%           \\ \hline
Domain                 & 50              & 71\%         & 70\%         & 30\%        & 18\%                     & 0\%                        & 2\%           \\ \hline
Consequence            & 42              & 65\%         & 86\%         & 14\%        & 7\%                     & 2\%                        & 5\%           \\ \hline
CPS                    & 31              & 94\%         & 94\%         & 6\%         & 0\%                      & 0\%                        & 0\%           \\ \hline
Perception             & 16              & 92\%         & 94\%         & 6\%         & 25\%                     & 6\%                        & 0\%           \\ \hline
Communication          & 0               & 100\%        & -          & -         & n=3                      & -                        & -           \\ \hline
Application            & 12              & 100\%        & 92\%         & 8\%         & 0\%                      & 0\%                        & 0\%           \\ \hline
\end{tabular}%
}
\end{table*}

Summarizing Table 5:
For the \textit{postmortem} fields,
 FAIL extracted 90\% or greater of the facts extracted by the analysts for the time, system, impacted organization, and impacts fields.
 However, FAIL only extracted 68\% of the facts extracted by the analysts for the non-software causes field.
 We conjecture that this performance is because news articles are more likely to focus on the observable impacts of incidents rather than the unobservable causes.
 This limitation of news articles makes it difficult to assess the non-software causes of an incident, even when done manually (40\% inter-rater agreement for our manual analysts).
For the multiple-choice (taxonomy) fields, FAIL and the analysts generally drew from the same facts: 
  overlapping was 90\%, missing was 10\%, additional and relevant was 18\%, additional and irrelevant was 6\%, and incorrect by the pipeline was 1\%.
\added{Notably, FAIL rarely introduced hallucinatory analysis, as we aimed (\cref{sec:DesignImplement}).
The average  proportion of incorrect facts introduced was 1\%, with a maximum of 5\%. 
The average proportion of irrelevant facts introduced was 8\%, where the facts themselves were not hallucinated, but the logical relationship argued by the LLM was unsound.}
Further details are in the table.

\subsection{Implementation details}

\textbf{Tech stack.}
Our pipeline implementation is ~3,500 lines of Python and 32 prompts.
We built the pipeline with a tech-stack of:
  Django application for interface,
  Postgres database for text data storage,
  Chroma DB as a vector database for RAG, and 
  LangChain to interact with the vector database.
The pipeline is periodically run to collect and analyze new incidents for the database using Celery.

\textbf{LLM selection.}
We used a SoTA LLM publicly available at the time of writing (2024): OpenAI's ChatGPT (gpt-3.5-turbo-0125)~\cite{ChatGPT}.
This LLM offers good performance at low cost~\cite{yaoTopNLPLanguage2023,gilardiChatGPTOutperformsCrowdWorkers2023a}.
It supports many human languages~\cite{HowManyLanguages}, so FAIL analyzes non-English articles.
\added{To reduce hallucination and to promote deterministic output, FAIL parameterized the LLM with a temperature of 0~\cite{zhangAutomatedRootCausing2024}.}

\textbf{Prompts.}
When working with an LLM, much of the implementation cost is in developing prompts.
As described in the design, FAIL uses LLM prompts in many steps.
FAIL uses a total of 32 prompts.
We applied best practices in prompt engineering, using iterative analysis and improvement as well as attempting a variety of strategies.
The prompts are available in our supplemental material.

\begin{tcolorbox} [width=\linewidth, colback=yellow!30!white, top=1pt, bottom=1pt, left=2pt, right=2pt]
\textbf{Requirement analysis:}
\begin{itemize}
    \item  R1-Scale: FAIL analyzes incidents over a 12-year span.
    \item  R2-Accuracy: FAIL's accuracy appears to be acceptable to fully or substantially automate news failure analysis, depending on the application.
    \item  R3-Cost: The cost of running FAIL to gather incidents over the span of 12 years was ~\$590, or ~\$0.25 per incident. 
\end{itemize}
\end{tcolorbox}

\section{Analyzing the FAIL DB} \label{sec:FailureAnalysis}

\JD{Signposting here}

\subsection{Research Questions}

Using FAIL we conducted a large-scale systematic study of recent software failures from the news.
We answer similar research questions as prior works that manually analyzed software failures~\cite{koThirtyYearsSoftware2014, rahmanIdentificationSourcesFailures2009, wongBeMoreFamiliar2017}.
Specifically, we ask:
{
\begin{RQList}
\item[RQ1:] What are the characteristics of the \textit{causes} of recent software failures? \JD{characteristics of the causes... seems unwieldy. How is that different from `what are the causes' ?}
\item[RQ2:] What are the characteristics of the \textit{impacts} of recent software failures? 
\JD{ditto}
\item[RQ3:] How do the \textit{causes} affect the \textit{impacts} of recent software failures? 
\end{RQList}
}
\DA{Should we change impacts to consequences?}

\JD{Paragraph explaining what we mean by `characteristics' and why we aren't stating causes/impacts directly (huge dataset, much more than prior work that could do it manually, automated analysis is left to future work, etc. --- give forwrad reference to Discusison section))}

\JD{Update these claims}
The contributions of our analysis are:
  (1) The last similar work was conducted in 2014, so we give an updated view as of 2024;
  and
  (2) Prior related work relies on manual analysis~\cite{rahmanIdentificationSourcesFailures2009, wongBeMoreFamiliar2017} or semi-automation~\cite{koThirtyYearsSoftware2014}, whereas ours uses a fully automated approach.

\subsection{Preliminary analysis of the database}
Using FAIL we collected \TotalIncidents{} incidents as reported in~\cref{table:FAILSteps}.
During \StepSearch{} of FAIL, 137,427 articles were found with our search query from Google News.
Out of these articles, FAIL successfully scraped 121,941 articles (88.73\%). 
During \StepClassifyFailure{} of FAIL, 6,553 articles (5.37\%) were classified to have reported on a software failure. 
During \StepClassifyAnalyzability{}, 4,184 articles (63.84\%) were classified to have met the criteria for sufficient information to conduct failure analysis. 
During \StepIndex{}, all of these articles were merged into 2,457 incidents. 

{
\begin{table}[ht]
\centering
    \caption{
    Number of articles for \StepSearch{} to \StepClassifyAnalyzability{} and the number of incidents for \StepIndex{} of the pipeline from 2010 to 2022.
    }
    \label{table:FAILSteps}
\small
\begin{tabular}{|c|p{1cm}|p{1cm}|c|c|c|c|}
\hline
\textbf{Year} & \textbf{\StepSearch{} (search)} & \textbf{\StepSearch{} (scraped)} & \textbf{\StepClassifyFailure{}} & \textbf{\StepClassifyAnalyzability{}} & \textbf{\StepIndex{}} \\
\hline
        2010 & 3999 & 3697 & 154 & 102 & 71 \\
        2011 & 6188 & 5709 & 266 & 188 & 120 \\
        2012 & 7247 & 6690 & 292 & 211 & 157 \\
        2013 & 7819 & 7292 & 317 & 222 & 161 \\
        2014 & 9557 & 8761 & 344 & 230 & 144 \\
        2015 & 10753 & 9888 & 425 & 290 & 182 \\
        2016 & 11745 & 10348 & 451 & 327 & 193 \\
        2017 & 9054 & 7273 & 483 & 284 & 151 \\
        2018 & 13089 & 11304 & 616 & 416 & 229 \\
        2019 & 14287 & 12856 & 766 & 541 & 271 \\
        2020 & 15358 & 13040 & 623 & 383 & 223 \\
        2021 & 13249 & 11799 & 869 & 539 & 269 \\
        2022 & 15082 & 13284 & 747 & 451 & 286 \\
        \hline
        \textbf{Total} & \textbf{137427} & \textbf{121941} & \textbf{6353} & \textbf{4184} & \textbf{2457} \\
\hline
\end{tabular}
\end{table}
}
\DA{Reformat this into Table 9}

\begin{figure}[!ht]
    \centering
    {\includegraphics[width=1\linewidth]{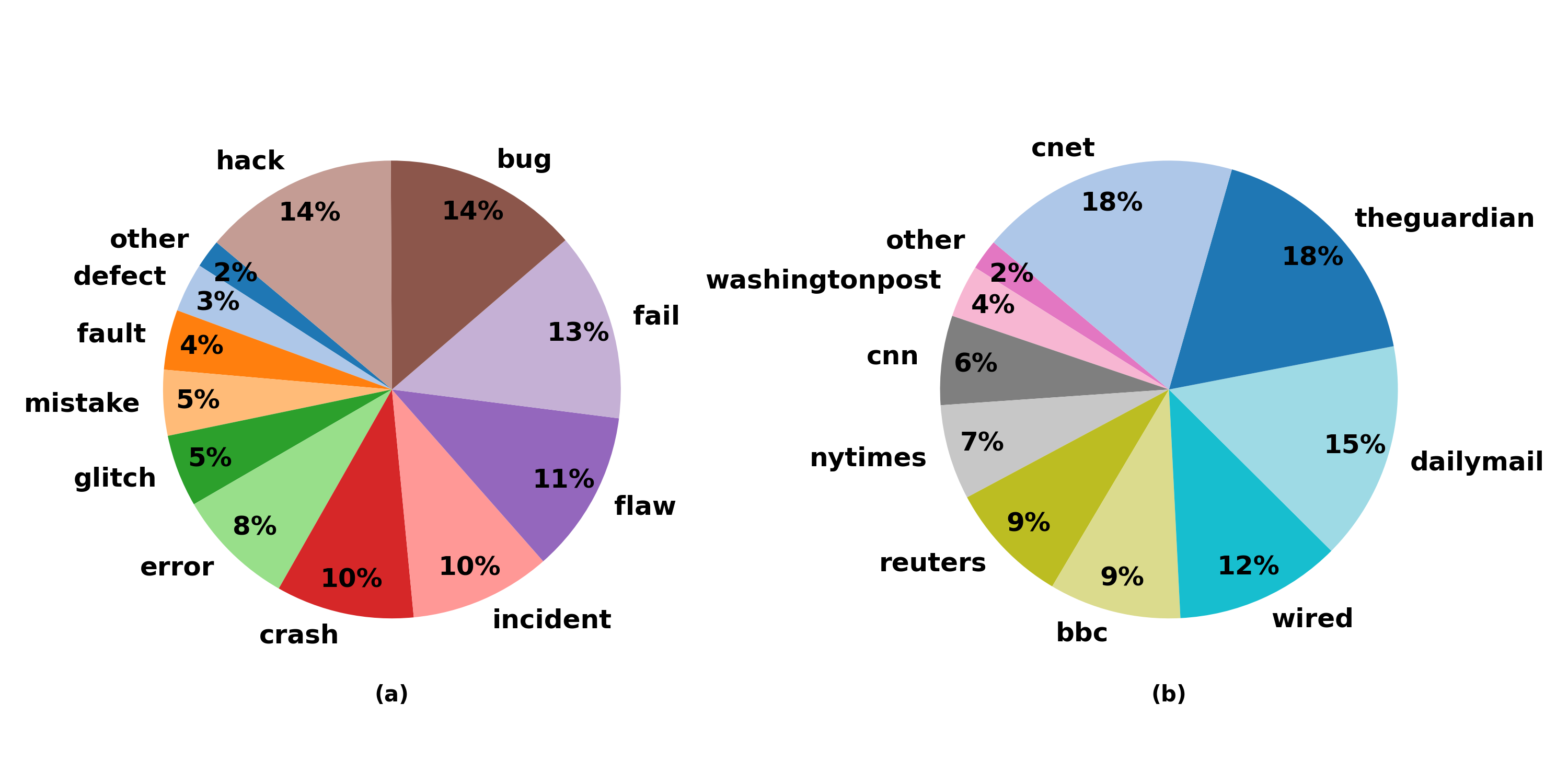}}
    \caption{
    (a) Incidents by keyword that found them in Step 1.
    (b) Distribution of news sources.
    }
    \vspace{-0.30cm}
    \label{fig:SourcesKeywordsPieChart}
\end{figure}

We report the distribution of keywords and the distribution of sources for the articles from the incidents in our database in~\cref{fig:SourcesKeywordsPieChart}.
The keywords that resulted in the most number of articles reporting on software failures are: "hack" (14\%) and "bug" (14\%), followed by "fail" (13\%), "flaw" (11\%), "incident" (10\%), "crash" (10\%), and error (8\%). 
The sources that published the most number of articles reporting on software failures are: CNET (18\%) and the Guardian (18\%), followed by Daily Mail (15\%), Wired (12\%), BBC (9\%), Reuters (9\%), New York Times (7\%), and CNN (6\%).

\begin{figure}[!ht]

    \centering
    {\includegraphics[width=0.8\linewidth]{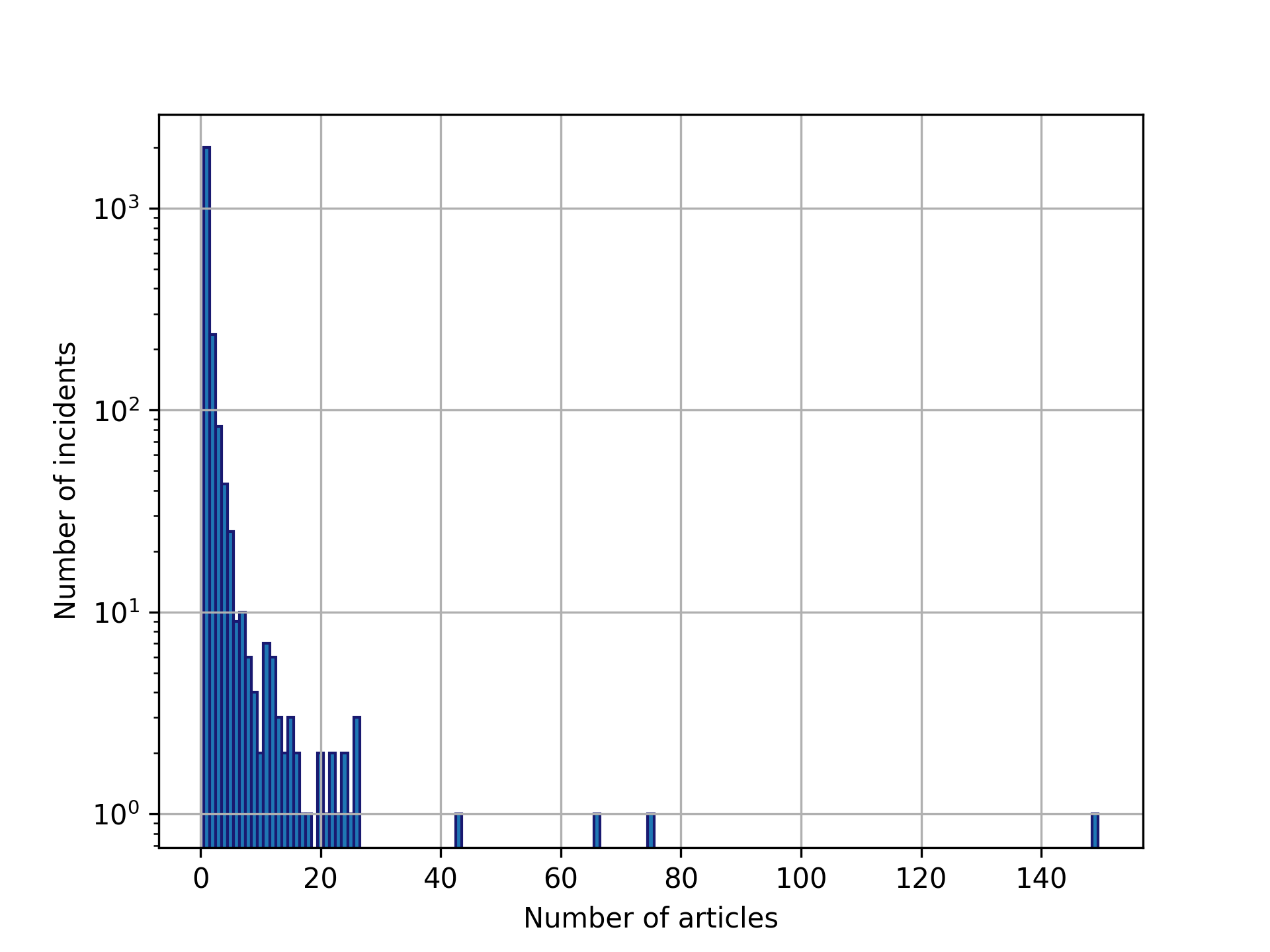}}
    \caption{
    Frequency of articles (x-axis) to incidents (y-axis, log scale).
    The most-covered incidents were the Boeing 737 MAX crashes, Sony PlayStation network attacks, and Tesla Autopilot failures.
    \JD{Comment on the fact that the typical case is indeed *more* than one article per incident, pretty sweet. It would be nice to have a deeper analysis indicating the number of facts per incident that come from 1 vs multiple articles.}
    }
    \vspace{-0.30cm}
    \label{fig:FrequencyArticlesIncidents}
\end{figure}

The number of incidents reported by the news sources and collected by FAIL increased over time as illustrated in~\cref{table:FAILSteps}.
As illustrated in~\cref{fig:FrequencyArticlesIncidents}, most of the incidents only contained one article: 81.28\%  (n=1997) of incidents only contained 1 article, 17.05\% (n=419) contained 2 to 10 articles, 1.1\% (n=27) contained 11 to 20 articles, 0.41\%  (n=10) contained 21 to 30 articles, and 0.16\% (n=4) contained more than 30 articles.
The four outliers were major incidents that received extensive news coverage.
Incident 1912 with 43 articles was about the solar winds security attack.
Incident 351 with 66 articles was about Tesla autopilot system failures that led to crashes. 
Incident 36 with 75 articles was primarily about a cyber-attack on Sony's PlayStation network, however many articles about cyber-attacks were incorrectly merged into this incident.
Incident 1396 with 150 articles was about the Boeing 737 MAX crashes.

\JD{Each of these needs to be structured with subsubsection for METHOD and RESULTS.}

\DA{For RQ2 and RQ3, talk about the high unknowns? For recurring, intent, communication}

\JD{TODO: Two stories}

For RQ1, RQ2, and RQ3, we highlight some of the interesting findings we observed from ~\cref{fig:TaxonomyDistributionSubplotsCauses} and ~\cref{fig:TaxonomyDistributionSubplotsImpacts}. 
\deleted{We illustrate the findings using the following randomly chosen incident from our database.
The summary is by the LLM, edited slightly for space.}

\added{We illustrate our findings using two incidents randomly chosen from the domains with the most incidents: information (\#2389) and transportation (\#2527).
\cref{table:ExampleFailureReports} presents excerpts from the LLM analysis of those incidents.} 

\begin{table*}
\caption{
\added{We illustrate two example failure reports from the FAIL database. 
We randomly chose an example from the two domains with the most incidents: information and transportation. 
We edited the LLM's prose for brevity and show only some of the analyzed fields.}
}
\label{table:ExampleFailureReports}
\centering
\begin{tabular}{|l|l|p{6.5cm}|p{6.5cm}|}
\cline{1-4}
\multirow{8}{*}[-22mm]{\centering \begin{sideways}\textit{Postmortem}\end{sideways}}  
& \textbf{Title}          & \textit{Parking Payment Machines Software Glitch Causes Financial Hardship in Worcester	 (Incident 2389: Articles 134147~\cite{timsWorcesterParkingGlitch2022}, 132785~\cite{WorcesterParkingMachine2022}) }                                                                                        & \textit{Security Flaw Exposes Social Security Numbers of Missouri Teachers (Incident 2527: Article 119973~\cite{technicaMissouriThreatensSue})}                                                                                                \\
\cline{2-4}
& \textbf{Summary}       & An erroneous software upgrade by the parking payment machines operated by Worcester City Council's parking contractor, Flowbird, caused duplicate payments to be debited from users' bank accounts. An estimated 1,500 drivers were overcharged. & A security flaw was identified in a website maintained by the Missouri Department of Elementary and Secondary Education (DESE). The flaw exposed the Social Security Numbers (SSN) of a 100,000 teachers and other school employees.                 \\
\cline{2-4}
& \textbf{Time}           & September, 2022                                                                                                                            & November, 2019                                                                                                                   \\
\cline{2-4}
& \textbf{System}         & Flowbird parking payment machines 
&1. Website maintained by DESE,
2. Educator-credentials checker system                                                                                                                                                                                                                                                                                                                                                                                                                                                                                                                                               \\
\cline{2-4}
& \textbf{Responsible Org} & 1. Worcester city council, 2. Parking contractor Flowbird 
& 1. DESE for maintaining the website,
2. The Governor blamed a journalist \& the St. Louis Post-Dispatch for reporting the vulnerability                                                                                                                                                                                                                                                                                                                                                                \\
\cline{2-4}
& \textbf{Impacted Org}    & Drivers in Worcester & Teachers and other school employees                                                                                                                                                                                                                                                                                      \\
\cline{2-4}
& \textbf{SE Causes}       & A bug introduced during a recent software upgrade & A security vulnerability: the data on DESE's website was encoded but not encrypted, making it relatively easy to decode                                                                                                                                                                                                                                                                                                                                                                                          \\
\cline{2-4}
& \textbf{Impacts}        & An estimated 1,500 people were overcharged & 100,000 school employees had their SSNs exposed                                                                                                                                                                                                                                                                                                                                            \\
\cline{1-4}
\multirow{5}{*}{\begin{sideways}\textit{Taxonomy}\end{sideways}}   & \textbf{Recurring}      & one organization, multiple organizations                                                                                      & one organization                                                                                                                                                                                                                                                                                                                                 \\
\cline{2-4}
& \textbf{Phase}          & design, operation                                                                                   & design, operation                                                                                                                                                                                                                                                                                                                                \\
\cline{2-4}
& \textbf{Domain}        & transportation, finance  & government, information, knowledge                                                                                                                                                                                                                                                                                                                                                                                                            \\
\cline{2-4}
& \textbf{Consequence}    & property & property                                                                                                                                                                                                                                                                                                                                                                                                                                   \\
\cline{2-4}
& \textbf{Cyb.-Phys. Sys.}            & Yes & No \\                                                                                                                                                                                                                                                                                \cline{1-4}                                                      
\end{tabular}%
\end{table*}

\deleted{FAIL's summary of incident 2389: The incident occurred in Worcester. It involved the parking payment machines operated by Worcester City Council's parking contractor, Flowbird. The failure started around mid-September when an erroneous software upgrade by Flowbird caused duplicate payments to be debited from individuals' bank accounts. This glitch led to an estimated 1,500 drivers being overcharged, with some individuals experiencing multiple unauthorized transactions, such as one individual being debited 122 times over three days, resulting in financial hardship for many affected drivers. The impact of the failure was significant: drivers out of pocket, some unable to pay bills, and one individual unable to afford a pending holiday. Responsible entities were Flowbird, the parking contractor, and Worcester City Council.
The council issued a breach notice to Flowbird and promised refunds.
Articles: [134147, 132785].
}

\subsection{\textbf{RQ1: } Characteristics of the \textit{causes} of recent software failures}

\begin{figure*}[!ht]
    \centering
    {\includegraphics[width=\linewidth]{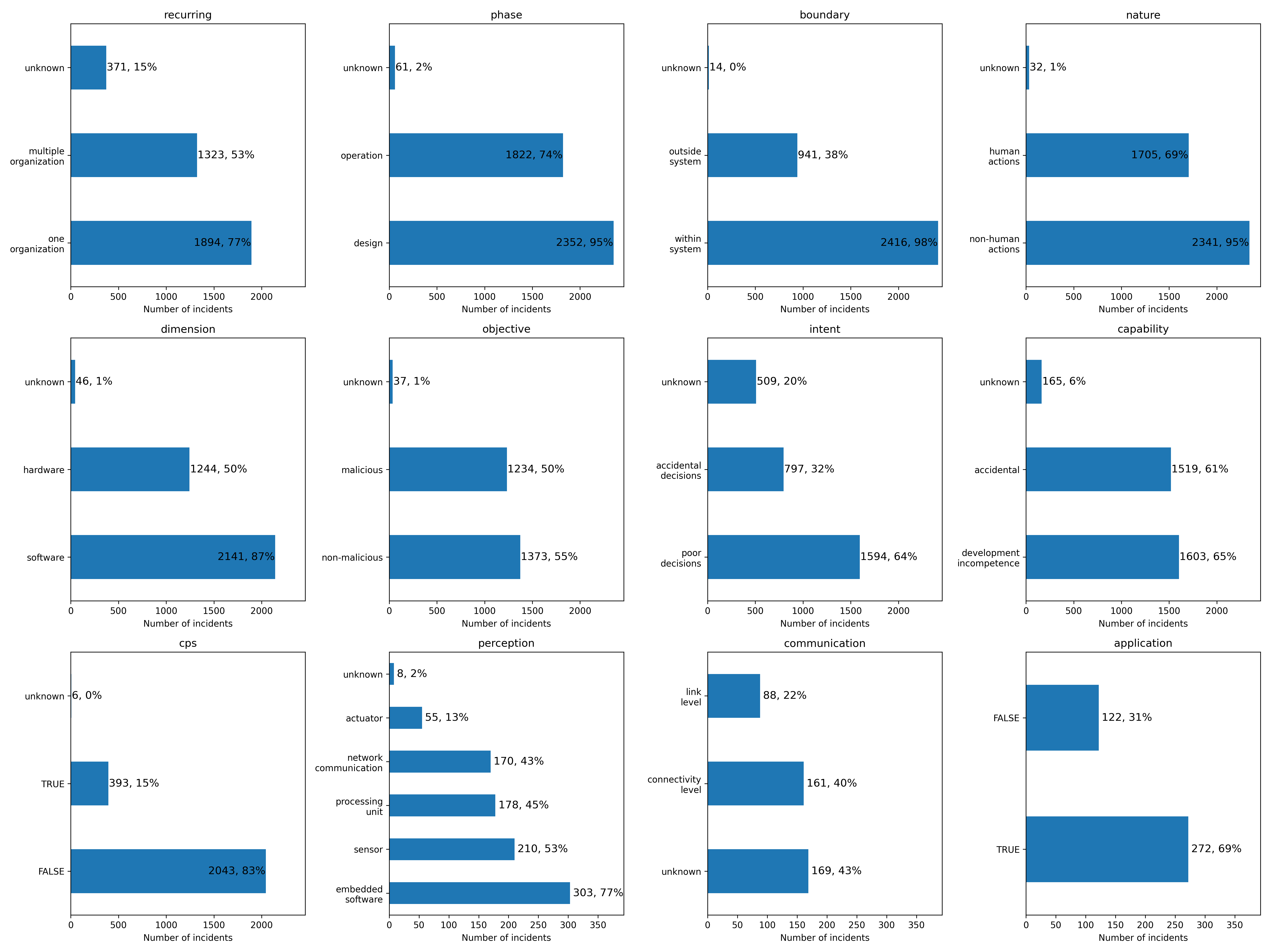}}
    \caption{
    Taxonimization of the characteristics of the \textit{causes} of recent software failures from our database.
    \JD{Increase font size.}
    }
    \vspace{-0.30cm}
    \label{fig:TaxonomyDistributionSubplotsCauses}
\end{figure*}

The taxonomization of the characteristics of the \textit{causes} of recent software failures analyzed by our pipeline is illustrated in~\cref{fig:TaxonomyDistributionSubplotsCauses}.

\textbf{Recurring: } 
\JD{This is the journalist's judgment of similarity --- whether they cite other stuff themselves, not based on our extraction}
\JD{Dharun says the LLM is probably wrong about this particular case?}
Most of the incidents (85\%) in our database were recurring incidents where a similar incident reoccurs across multiple organizations (53\%) and/or recurring at one organization (77\%).
Incident \#2389 was a recurring incident within the organizations involved and had similar incidents occur at other organization from the same developer, where there were recurring issues with the payment processing for parking. 
\textbf{Phase: } 
Most of the failures were due to contributing factors introduced during system design (95\%), with 75\% of the failures due to contributing factors introduced during operation.
Incident 2389 occurred due to factors originating from the design and operation phases, since it was because of a software glitch from a software update. 
\textbf{Boundary: } 
Most of the failures in our database were due to contributing factors originating within the system (98\%), with 38\% of the failures due to contributing factors originating outside the system.
Incident \#2389 was caused by a software glitch within the system, and was exacerbated by continued unauthorized debits from individuals' bank accounts after being notified of the issue by the city council.
\textbf{Nature: } 
Most of the failures in our database were due to contributing factors that did not involve human-actions (95\%), with 69\% of the failures due to contributing factors involving human actions.
Beyond the software glitch, incident \#2389 was exacerbated by the poor response by the employees of the company.
\textbf{Dimension: } 
Most of the failures in our database were due to contributing factors that originate in software (87\%), with 50\% of the failures due to contributing factors originating in hardware.
Incident \#2389 occurred due to only software issues, with no reports of hardware issues.
\textbf{Objective: } 
Roughly an equal amount of the failures in our database were due to contributing factors introduced by human(s) with intent to harm a system (50\%), with 55\% of the failures due to contributing factors introduced without intent to harm a system.
Incident \#2389 did not have any malicious factors.
\textbf{Intent: } 
A majority of the failures in our database were due to contributing factors introduced by poor decisions (64\%), with 32\% of the failures due to contributing factors introduced by accidental decisions.  
Incident \#2389 was attributed to poor and accidental decisions by the developer. 
\textbf{Capability: } 
Roughly an equal amount of the failures in our database were due to contributing factors introduced accidentally (61\%) and introduced by development incompetence (65\%). 
Incident \#2389 was due to accidental factors as well as developmental negligence. 

\textbf{Cyber-Physical System: }
15\% of the software failures in our database were in Cyber-Physical Systems (CPS).
Incident \#2389 was due to a failure in a system for parking payment machines, which is a cyber-physical system.
\textbf{Perception: }
Most failures in the perception layer of CPS systems in our database were due to contributing factors introduced by embedded software (77\%), followed by sensors (53\%), processing unit (45\%), network communication (43\%), and actuators (13\%).
Incident \#2389 could have occurred due to factors involving the processing unit and the embedded software.
\textbf{Communication: }
Most failures in the communication layer of CPS systems in our database were due to contributing factors introduced by connectivity level (40\%), followed by link level (22\%).
It is unclear whether incident \#2389 involved the communication layer. 
\textbf{Application: }
69\% of the failures in the CPS systems in our database are due to contributing factors introduced by the application layer. 
Incident \#2389 involved factors in the application layer.

\subsection{\textbf{RQ2: } Characteristics of the \textit{impacts} of recent software failures}
\begin{figure*}[!ht]
    \centering
    {\includegraphics[width=\linewidth]{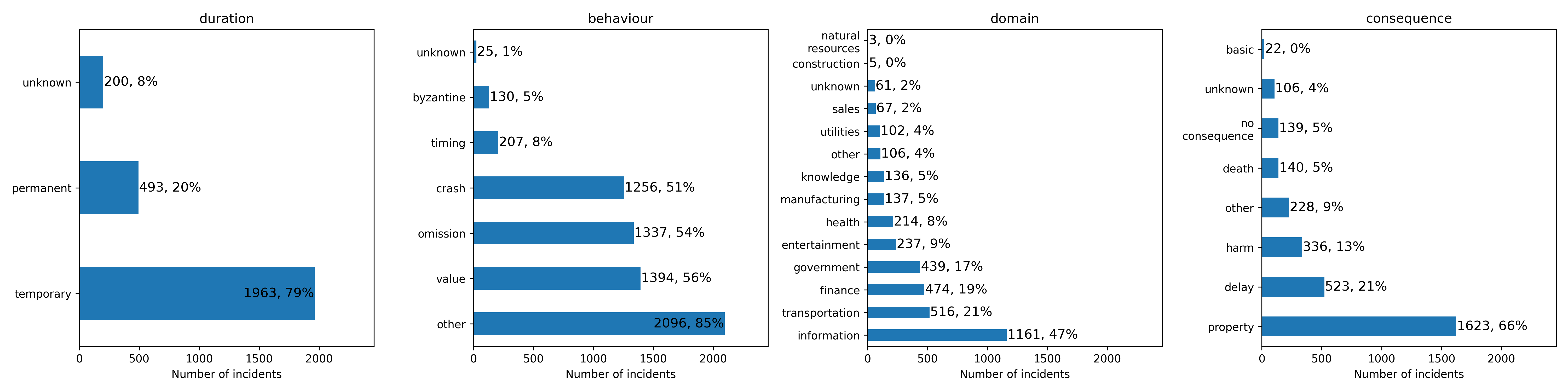}}
    \caption{
    Taxonimization of the characteristics of the \textit{impacts} of recent software failures from our database.
    }
    \vspace{-0.30cm}
    \label{fig:TaxonomyDistributionSubplotsImpacts}
\end{figure*}

The taxonimization of the characteristics of the \textit{impacts} of recent software failures analyzed by our pipeline is illustrated in~\cref{fig:TaxonomyDistributionSubplotsImpacts}.

\textbf{Duration: } 
Most of the failures in our database were due to contributing factors that were temporary (79\%), with 20\% of the failures due to contributing factors that were permanent. 
Incident \#2389 was caused by temporary factors. 
\textbf{Behaviour: } 
Roughly an equal amount of the failures in our database were due to a system crashing (51\%), a system omitting to perform its intended function (54\%), and a system performing its function incorrectly (56\%).
Notably, 85\% of the failures in our database were due to a system behaving in a way not described in the taxonomy~\cite{avizienisBasicConceptsTaxonomy2004a}.
Incident \#2389 was due to the system crashing, omitting to charge users at certain instances, charging users at incorrect times, and charging users incorrect values. 
\textbf{Domain: } 
The software failures in our database most commonly occurred in systems supporting industries related to information (47\%), followed by transportation (21\%), finance (18\%), and government (17\%).
The software failures also occurred in systems supporting industries related to entertainment (9\%), health (8\%), manufacturing (5\%), knowledge (5\%), utilities (4\%), and sales (2\%).
Incident \#2389 involved transportation and financial application domains. 
\textbf{Consequence: } 
The most common consequence of the software failures in our database was impact to peoples' (material goods, money, or data) property (66\%), followed by delay to peoples' activity (21\%), physical harm to people (13\%), and people losing their lives (5\%).
As a result of incident \#2389, customers' money was impacted.  

\subsection{\textbf{RQ3: } Influence of the \textit{causes} on the \textit{impacts} of recent software failures}

\textbf{Recurring}:
Similar incidents reoccur across multiple organizations most in the information (27\%), finance (13\%), government (11\%), and transportation (9\%) domains.
Most cyber-physical system failures had similar recurrences (14\%).
\textbf{Objective}:
Incidents with malicious objectives mostly occurred in the information (29\%), finance (13\%), and government (11\%) domains. 
Incidents with malicious objectives mostly only impacted property (40\%).
About half of the failures in cyber-physical systems were due to malicious intent (9\%), which mostly impacted property (6\%) and caused some harm (2\%).
\textbf{Cyber-physical system: }
Most cyber-physical system failures occurred in the transportation domain (7\%), followed by the information (4\%), government (3\%), and health (2\%) domains. 
These failures primarily impacted property (9\%) and led to harm (4\%).

\section{Discussion and Future work} \label{sec:Discussion}

\subsection{Recurring incidents}
Our results for RQ1 indicate high recurrences of similar failures: within organizations and across organizations. 
This implies that the software engineering community might benefit from learning from past failures to prevent similar future ones. 
To reduce recurring incidents within organizations, we first need to study the current practices utilized by practitioners to learn from failures within organizations and their effectiveness. 
To reduce recurring incidents across organizations, we suggest the need for the software engineering community to learn from inter-organizational failures using resources such as the Risks Digest~\cite{neumannRISKSDigest} or our database.
Specifically, based on our results for RQ3, the information, finance, government, and transportation domains had a high rate of similar incidents recurring and may benefit from an inter-organizational failure database within each domain.

\subsection{Change in the consequences of software failures over time} \label{sec:Discussion-Consequences}
The consequences of software failures have evolved in severity over time. 
In Ko \etal's analysis of software failures from 1980 to 2012, 50\% of the failures had no consequences, 40.4\% delayed activities, 12.5\% impacted property, 1.2\% led to death, 1\% led to harm, and 0.7\% impacted basic needs. 
Whereas, based on our results for RQ2, for software failures from 2010 to 2022, we find that only 5\% had no consequences, with increase in failures that impacted property, led to harm, and led to death. 
This indicates that the severity of the consequences of software failures have increased over the past decade.
This may be due to the increased use of software in modern society.
The increase in the severity of the consequences of software failures, indicates the need to develop safer software. 
As such, policymakers, practitioners, and academics should study the causes of software failures and their consequences to prioritize preventing high severity software failures.

\subsection{Cybersecurity failures}
Our results for RQ1 indicate that half of the recent software failures involved malicious intent.
These failures were more common in the information, finance, and government domains (RQ3). 
Results were similar for the failures in cyber-physical systems.
This data motivates continued investment in improving the cybersecurity of our digital and physical systems. 

\subsection{Applications: Policy, Engineering, Research}

We invite the policy-making, engineering, and academic communities to explore insights and applications that could be built from our database of software failures.
The use of software in the modern world is ubiquitous, and based on~\cref{sec:Discussion-Consequences}, the consequences of their failures have become more severe. 
To enable the safe development and implementation of software, we advocate exploring our database to learn how software failures occur, what their impacts are, and how to prevent them. 
This insight could inform policies around software:
what are the ways in which people are affected by software failures?
what policies, regulations, and standards could help prevent such failures?
This insight could also inform software engineering: 
what are the common causes of software failures and how can we prevent them?
what context are software systems used in and how do they invoke failures?
This insight could also inform academics: 
it could guide research on improving software reliability and developing new methods for failure prevention.
Additionally, given the high occurrences of incidents with malicious intent, researchers could map the failures in the database with the Common Weakness Enumeration (CWE) categories to identify the common weaknesses in real world security failures.

\subsection{Extension to FAIL}
We propose extensions to FAIL to leverage failure knowledge from our database of software failures. 
First, our database contains open-ended fields with rich failure data such as the: causes, impacts, and  mitigations for failure.
LLMs can be used to conduct thematic analysis~\cite{depaoliPerformingInductiveThematic2023, daiLLMintheloopLeveragingLarge2023} on this data to identify failure trends. 
Second, we propose a failure-aware chat-bot. 
An LLM-based chatbot could be developed with our database to aid engineers during the SDLC.
For example, grounded by past failure knowledge, a failure-aware chatbot could: aid in refining requirements to improve resiliency of a system, in the planning phase, or offer design rationales to aid with design decisions, in the design phase, or write test cases to look out for past failure causes, in the testing phase, or suggest mitigation strategies for incidents, in the maintenance phase.

\added{We expect that FAIL will need to evolve to keep pace with external changes.
In terms of its primary components, FAIL can be updated to take advantage of any improvements in news aggregators, as well as new LLMs such as LLama3. 
Conceptually, the main need for evolution is driven by engineering technologies.
Just like any other failure analysis technique, as computing technology changes, FAIL will need to be updated.
Each step of FAIL involves some definition of ``software failure'', and as technology changes over time, FAIL will need to be adapted by updating these definitions.
For example, microservice and serverless architectures made “grey failures” more common~\cite{huang2017gray}, and the advent of AI-driven systems has also introduced new failure modes.
New taxonomies are being introduced for such failures, \eg the CSET taxonomy for AI incidents~\cite{csetTaxonomy}.
The FAIL database itself can then be kept up to date through periodic runs on new tranches of news data.
}

\section{Threats to Validity} \label{sec:Threats}

We discuss three types of threats to validity~\cite{wohlin2012experimentation}, focusing on substantive threats that might influence our findings~\cite{verdecchia2023threats}.

\textbf{Construct Threats} are potential limitations of how we operationalized concepts.
In this work we relied on existing constructs (\cref{sec:background}):
  failures and incidents are well documented concepts,
  and
  we applied existing taxonomies from prior work.
We scoped our definition of news articles to mainstream media sources such as CNN and Wired, which may reduce the level of technical detail available.

\textbf{Internal threats} are those that affect cause-effect relationships.
This paper propagates the failure cause-effect relationships as indicated in the articles selected by FAIL.
In addition, FAIL relies on a Large Language Model (LLM) which may hallucinate and introduce incorrect cause-effect relationships.
\added{To mitigate, we carefully engineered the prompts used by FAIL and parameterized the LLM with a temperature of 0 (detailed in~\cref{sec:Component Development and Validation}).}
In our evaluation, we found this was relatively rare for most fields.
Additionally, we acknowledge that our evaluation of FAIL was based on comparison to human analysts, and that the analysts may themselves have erred in their interpretation of the studied articles.
To mitigate, we used two analysts and found reasonably high inter-rater agreement.

\textbf{External threats} may impact generalizability.
Any application of failure data drawn from the news media must acknowledge the potential biases of that media~\cite{mullainathanMediaBias2002}.
A publication's editors may sway which events are covered and with what frequency.
For example, the FAIL database is likely biased toward more large-scale, impactful, and ``interesting'' incidents.
The FAIL database is also likely geographically biased.
It does include articles from multiple languages, but as Ko \etal noted~\cite{koThirtyYearsSoftware2014}, failure coverage tends to focus on issues in the United States and Europe.

\section{Conclusion} \label{sec:Conclusion}

\GKT{As currently written, this does not have a pop. What specific results make the reader say that we're the best.}
\JD{Dharun, PTAL and refine as needed}
\GKT{It needs work.}

In this work, we presented the concept of \ul{F}ailure \ul{A}nalysis \ul{I}nvestigation with an \ul{L}LM. \GKT{Seriously, we might need a macro for this!}
We materialized this concept in the FAIL system, which applies OpenAI's GPT-3.5 large language model to news articles from 11 sources including The New York Times, the BBC, and Wired Magazine.
Where prior work relied on manual analysis of such articles, FAIL demonstrates the novel capability of automated analysis.
We learned that the current generation of large language models is capable of identifying news articles that describe failures, and analyzing them according to structured taxonomies.
FAIL will thus enable researchers, practitioners, and policymakers to keep up \GKT{don't laugh: succeed} with the pace of public software failure data.

\section{Data Availability} \label{sec:DataAvailability}
The code, prompts, manual analysis, and  evaluation for FAIL, as well as a publicly viewable version of our database is available at: \url{https://zenodo.org/doi/10.5281/zenodo.13761264}. 

\ifBLINDED
\else
\section*{Acknowledgments}

We thank G.K. Thiruvathukal for his thorough feedback on the manuscript,
and
P.V. Patil for help typesetting the bibliography.

\fi

\raggedbottom
\pagebreak

\balance
\bibliographystyle{bib/ACM-Reference-Format}
\bibliography{bib/compressed-bib,bib/short}

\end{document}